\documentclass[onecolumn]{article}

\usepackage[dvipdfm]{graphicx}
\usepackage{amsmath,amssymb,bm,cases,subfigure,arydshln}
\usepackage{cite}
\title{\LARGE \bf
Turing Instability in Reaction-Diffusion Systems with a Single Diffuser: 
Characterization Based on Root Locus
\footnote{\copyright 2013 IEEE. Personal use of this material is
permitted. Permission from IEEE must be obtained for all other uses, in
any current or future media, including reprinting/republishing this
material for advertising or promotional purposes, creating new
collective works, for resale or redistribution to servers or lists, or
reuse of any copyrighted component of this work in other works. 
}
}
\date{}
\author{Hiroki Miyazako, Yutaka Hori and Shinji Hara%
\thanks{H. Miyazako, Y. Hori and S. Hara are with the Department of Information Physics and
        Computing, The University of Tokyo, Tokyo 113-8656, Japan.
   {\tt \small \{Hiroki\_Miyazako, Yutaka\_Hori, Shinji\_Hara\}@ipc.i.u-tokyo.ac.jp}}%
}
 \evensidemargin \oddsidemargin

\begin{document}
\maketitle

\thispagestyle{empty}
\pagestyle{empty}

\begin{abstract}
Cooperative behaviors arising from bacterial cell-to-cell communication 
 can be modeled by reaction-diffusion equations having only a single diffusible component.
This paper presents the following three contributions for the systematic 
 analysis of Turing instability in such reaction-diffusion systems. 
(i) We first introduce a unified framework to formulate the
 reaction-diffusion system as an interconnected multi-agent dynamical system. 
(ii) Then, we mathematically classify biologically plausible and implausible Turing
 instabilities and characterize them by the root locus of each agent's
 dynamics, or the local reaction dynamics.
(iii) Using this characterization, we derive analytic conditions for
 biologically plausible Turing instability, which provide useful guidance 
 for the design and the analysis of biological networks.
These results are demonstrated on an extended Gray-Scott model with a single diffuser.
\end{abstract}

\section{Introduction}
In biological compartments, spatial diffusion of molecules can lead to 
an ordered spatial pattern of chemical concentrations (see \cite{Kondo2010,Loose2011}
for examples).
The dynamical model of such diffusion-driven pattern formation, or
Turing pattern formation, was first introduced by Turing \cite{Turing1952} as a reaction-diffusion system. 
Then, intensive efforts during the past half-century have revealed
essential mechanisms of diffusion-driven instability, or Turing instability, such as 
the activator-inhibitor theory \cite{Gierer1972, Koch1994, Edelstein-Keshet2005}.

\par
\smallskip
In recent synthetic biology, researchers have attempted to design 
gene regulatory networks that result in spatially patterned gene expression over a
population of bacterial cells \cite{Basu2004, Basu2005}. 
One of the ideas to realize such a biological circuit is to utilize a 
cell-to-cell communication mechanism mediated by a diffusible molecule called an
autoinducer \cite{Miller2001}. 
We can then expect that the diffusion of the autoinducer drives a certain
spatial mode and generates spatial patterns of gene expression.
In contrast to the classical ones, 
 the reaction-diffusion model for such system has a
distinctive feature that there is only one diffusible molecule, or the
 autoinducer, in the system. 
Thus, a novel theoretical framework is desirable to
 systematically study reaction-diffusion systems having only one
 diffusible component.

\par
\smallskip
In particular, there still remains an important fundamental question
that whether Turing patterns can be generated in reaction-diffusion
systems with a single diffuser. 
This question was partly tackled in the recent theoretical study 
\cite{Hsia2012}, where a gene regulatory network showing spatial patterns
was proposed.
Our study, however, has verified that the patterns emerging from the
examples in \cite{Hsia2012} result in extremely fine spatial
patterns dominated by infinitely large spatial frequency, which is
biologically implausible. 
Hence, it is desirable to theoretically characterize biologically
plausible Turing instability and to develop a systematic way to 
explore the conditions for such instability.

\par
\smallskip
This paper presents the following three contributions on
reaction-diffusion systems with a single diffuser. 
\begin{itemize}
\item[(i)] We propose a unified framework to systematically formulate
 the reaction-diffusion system as an interconnected multi-agent
			dynamical system.
\item[(ii)] We then characterize biologically plausible and implausible Turing
instabilities by the root locus of the local reaction dynamics.
\item[(iii)] Using this characterization, we derive analytic conditions for biologically plausible Turing
instability. 
\end{itemize}
In particular, our analysis proves that it is possible to generate
biologically plausible patterns only with a single diffusible
molecule. These results provide useful guidance for the design of synthetic biological circuits.

\par
\smallskip
The organization of this paper is as follows. 
In the next section, we introduce the reaction-diffusion system
considered in this paper and formulate it as a multi-agent dynamical system.
In Section III, we mathematically define biologically
plausible and implausible Turing instabilities. 
Then, in Section IV, these instabilities are
characterized by the root locus.
In Section V, we derive analytic conditions for the biologically plausible
Turing instability. Then, Section VI is devoted to the demonstration of our result 
on an extended Gray-Scott model \cite{Vastano1988}.
Finally, Section VII concludes this paper.

\par
\smallskip
The following notations are used throughout this paper.
$\mathbb{C}_+ := \{s \in \mathbb{C}~|~\mathrm{Re}[s] > 0\}$ and 
$\mathbb{C}_{0+} := \{s \in \mathbb{C}~|~\mathrm{Re}[s] \ge 0\}$
. 
$\mathbb{R}_{\ge0} := \{c \in \mathbb{R}~|~c \ge 0\}$.
$\mathbb{N} := \{1,2,3,\cdots\}$.

\medskip

\section{Control Theoretic Formulation of Reaction Diffusion Systems}
In this section, we first introduce reaction-diffusion systems with a single
diffuser and show that the systems can be viewed as multi-agent
dynamical systems.

\subsection{Reaction-Diffusion systems with a single diffuser}
We consider a set of chemical reactions consisting of $n$ molecular
species, $\mathcal{M}_1, \mathcal{M}_2, \cdots, \mathcal{M}_n$, in one
dimensional space $\Omega := [0,L]$. 
Let $x_i(\xi, t)$ denote the concentration of
$\mathcal{M}_i~(i=1,2,\cdots,n)$ at position $\xi \in \Omega$ and at time
$t$, and define ${\bm x}(\xi, t) := [x_1(\xi,t), x_2(\xi,t), \cdots,
x_n(\xi,t)]^T$ \footnote{
In what follows, we may denote ${\bm x}(\xi, t)$ by ${\bm x}$ to avoid notational clutter.
}.
In this paper, we consider the case where only $\mathcal{M}_n$ can diffuse in the spatial domain.
The dynamics of the reactions and diffusion is then given by 
\begin{align}
\frac{\partial {\bm x}}{\partial t} = f({\bm x}) + D {\bm \nabla}^2 {\bm
 x}, 
\label{nonlinear-eq}
\end{align}
where the nonlinearity $f(\cdot)$ is a Lipschitz continuous function representing the dynamics of local
reactions, and $D := \mathrm{diag}(0,\cdots,0,\mu) \in \mathbb{R}_{\ge 0}^{n
\times n}$ is the matrix with a diffusion coefficient $\mu$.
It should be noted that only the $(n,n)$-th entry is non-zero, since
only $\mathcal{M}_n$ can diffuse. 
Throughout this paper, we assume the Neumann boundary condition as follows.
\begin{align}
\frac{\partial {\bm x}}{\partial \xi}(0,t) = 0,~~
\frac{\partial {\bm x}}{\partial \xi}(L,t) = 0.
\end{align}

\par
\smallskip
We here introduce a linearized model of (\ref{nonlinear-eq}) to analyze
Turing patterns based on local stability analysis in the following sections.
Let $\bar{{\bm x}} \in \mathbb{R}_{\ge 0}^{n}$ denote a spatially homogeneous equilibrium of
(\ref{nonlinear-eq}). The linearized system around $\bar{{\bm x}}$ is then obtained as 
\begin{align}
\frac{\partial \tilde{\bm x}}{\partial t} = A \tilde{\bm {x}} + D
 {\bm \nabla}^2 \tilde{{\bm x}},
\label{linear-eq}
\end{align}
where $A \in \mathbb{R}^{n \times n}$ is the Jacobian of $f(\cdot)$ evaluated at $\bar{{\bm x}}$, 
and $\tilde{\bm x}$ %
is defined by $\tilde{{\bm x}} := {\bm x} - \bar{{\bm x}}$.

\subsection{Formulation as a multi-agent dynamical system}
\begin{figure}[tb]
\centering
\includegraphics[clip,width=6cm]{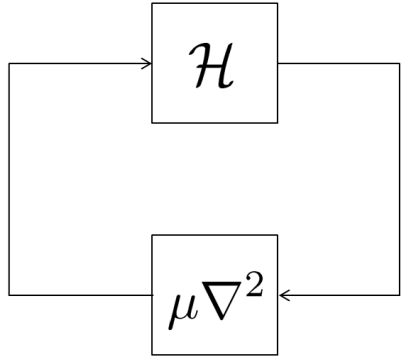}
\caption{Block diagram of the system (\ref{statespace-eq}). }
\label{overall-fig}
\end{figure}

\par
\smallskip
The linearized model (\ref{linear-eq}) can be equivalently written as 
\begin{align}
\begin{array}{cc}
\displaystyle \frac{\partial {\tilde{{\bm x}}}}{\partial t}=A \tilde{{\bm
 x}} + {\bm e}_n u\\
y = {\bm e}_n^T \tilde{{\bm x}},
\end{array}
\label{statespace-eq}
\end{align}
where $u = \mu \nabla^2 y$
 and ${\bm e}_n := [0,\cdots, 0,1]^T \in
\mathbb{R}^{n}$.
The equations (\ref{statespace-eq}) imply that for each fixed position $\xi$, the dynamics of
 local reactions, which we denote by $h(s)$, can be modeled as a SISO linear time-invariant system with
 the input $u(\xi, t)$ and the output $y(\xi, t)$. 
Specifically, $h(s)$ can be written as 
\begin{align}
h(s) := {\bm e}_n^T (sI_n - A)^{-1} {\bm e}_n~ (=:n(s)/d(s)), 
\end{align}
where we define $n(s)$ and $d(s)$ as the numerator and the denominator 
of $h(s)$, respectively.
The reaction-diffusion system (\ref{statespace-eq}) can then be 
interpreted as the feedback system illustrated in
Fig. \ref{overall-fig}, 
where 
$\mathcal{I}$ is an identity operator, and 
$\nabla^2 = \partial^2/\partial \xi^2.$
Note that this system can be viewed as a multi-agent dynamical system, 
where the homogeneous dynamical agents $h(s)$ are coupled with
the nearest neighbor agents by the Laplace operator, and 
the stability analysis of such class of systems was studied 
in \cite{Tanaka2009, Hara2013}.

\begin{figure}[tb]
\centering
\includegraphics[clip,width=10.5cm]{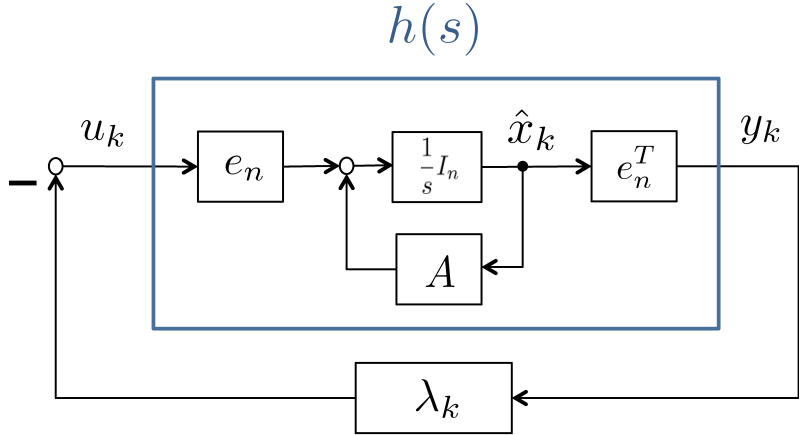}
\caption{Block diagram of the system $\Sigma_k~(k=0,1,2,\cdots)$.}
\label{diagonal-fig}
\end{figure}

\par
\smallskip
The infinitely large-scale multi-agent system can be decomposed into 
small subsystems by diagonalizing the Laplace operator $\nabla^2$ as 
$\mathcal{F}\nabla^2 \mathcal{F}^{-1} = \mathrm{diag}[h(s), h(s), \cdots]$ with the Fourier transform
operator $\mathcal{F}$.
The eigenvalues $\lambda_k$ and the associated eigenfunctions
$\phi_k(\xi)$ of $- \mu \nabla^2$ are
specifically obtained as 
\begin{align}
\lambda_k \!:=\! \mu \left(\frac{k \pi}{L} \right)^2,~\phi_k(\xi) \!:=\!
 \cos \left(\frac{k\pi \xi}{L} \right)
\label{lambda-def}
\end{align}
for $k=0,1,2,\cdots$.
Consequently, the closed-loop system in Fig. \ref{overall-fig} can be
decomposed into the subsystems $\Sigma_k~(k=0,1,2,\cdots)$ depicted in
Fig. \ref{diagonal-fig}. %

\par
\smallskip
It should be noticed that the dynamics of each subsystem $\Sigma_k$ 
represents that of each spatial mode $\phi_k(\xi)$. 
In fact, it follows that 
\begin{align}
\tilde{{\bm x}}(\xi, t) = \sum_{k=0}^{\infty} \hat{{\bm x}}_k(t) \cos\left(\frac{k \pi \xi}{L}\right), 
\end{align}
where $\hat{{\bm x}}_k$ is the state of the subsystem $\Sigma_k$ as depicted in Fig. \ref{diagonal-fig}.
In particular, the spatial pattern formation is expected when the growth rate 
of a non-zero spatial mode $\phi_k(\xi)~(k=1,2,\cdots)$ is positive. 
Therefore, the analysis of pattern formation in reaction-diffusion
systems reduces to the stability analysis of 
the small subsystems $\Sigma_k$. %

\par
\smallskip
Let the characteristic polynomial of the 
closed-loop system composed of $h(s)$ and a gain $\lambda$ be denoted by 
\begin{align}
p(\lambda, s) := d(s) + \lambda n(s).
\end{align}
The characteristic polynomial of each subsystem $\Sigma_k$ is then
 written as $p_k(s) := p(\lambda_k, s)$. 
Thus, a subsystem $\Sigma_k$ is asymptotically stable, if and only if
 $p_k(s)$ is Hurwitz. 
Moreover, we can also see that the linear infinite dimensional system (\ref{linear-eq}) is exponentially stable 
if and only if $p_k(s)$ is Hurwitz for all $k=0,1,2,\cdots$
\footnote{We can prove this since the Laplace operator is a Riesz
spectral operator, though the proof of the sufficiency requires careful treatment in general
infinite dimensional linear systems (see Section 5 of \cite{Curtain1995} for
details).}. 
\medskip

\section{Definitions of Biologically Plausible and Implausible Turing Instability}
In this section, we first introduce the definition of Turing
instability, by which the reaction-diffusion system exhibits spatial
patterns. Then, the Turing instability is classified into two types 
from a  viewpoint of biological plausibility. 

\medskip
\noindent
{\bf Definition 1.~}
{\it 
The reaction-diffusion system with a single diffuser modeled by
(\ref{nonlinear-eq}) is Turing unstable around an equilibrium $\bar{{\bm x}}$, 
if

\smallskip
\noindent
(i) $p_0(s) = d(s) \neq 0${\rm ;} $\forall s \in \mathbb{C}_{0+}$ and \\
(ii) $p_k(s) = 0${\rm ;} $\exists s \in \mathbb{C}_{0+}$ and 
 $\exists k \in \mathbb{N} \cup \{\infty\}$.

\smallskip
\noindent
That is, (i) all the poles of $\Sigma_0$ lie in the open left half-plane, {\it
i.e.,} $A$ is Hurwitz, and (ii) at least one pole of $\Sigma_k$ lies 
in the close right half-plane for some $k = 1,2,\cdots$.
}%

\par
\medskip
We can see that at least one non-zero spatial mode has a positive growth
rate around the equilibrium when the system (\ref{nonlinear-eq}) is Turing
unstable. 
Thus, we expect that the reaction-diffusion system exhibits a spatial
pattern after a small perturbation to the homogeneous equilibrium $\bar{\bm x}$. 
It should be noticed that the stability of $\Sigma_0$ implies that 
the dynamics of local reactions $h(s)$ is stable, {\it i.e.,} $A$ is Hurwitz, 
thus Turing instability is exclusively driven by the diffusion of the molecule $\mathcal{M}_n$.

\par
\smallskip
When there are multiple unstable subsystems $\Sigma_k$, the spatial
profile of the pattern depends on the growth rate of the subsystems' outputs.
More specifically, the spatial mode corresponding to the most unstable
subsystem tends to be dominant as $t \rightarrow \infty$.
Thus, the rightmost pole of the closed-loop system is of
particular interest in our study.

\medskip
\noindent
{\bf Definition 2.~}
{\it 
A pole $\sigma$ of $\Sigma_k$ is called 
a dominant closed-loop pole, if (i) $\mathrm{Re}[\sigma] \ge 0$ and 
(ii) $\sigma$ has the largest real part among all the poles of 
$\Sigma_k~(k=0,1,2\cdots)$. 
More precisely, the set of all dominant poles is defined by
\begin{align}
\Pi := \{&s_0 \in \mathbb{C}_{0+}~|~p_k(s_0) =
 0; \exists k \in \mathbb{N}\cup\{\infty \}, \mathrm{and} \notag \\
&p_k(s \!+\! \mathrm{Re}[s_0]) \neq
 0; k \in
 \mathbb{N}\!\cup\!\{\infty\}~\mathrm{and}~\forall s \!\in\!
 \mathbb{C}_+\}.
\notag
\end{align}
}%

\par
When the dominant pole is given by a subsystem $\Sigma_k$, the output of $\Sigma_k$ has the largest growth rate 
, thus the $k$-th spatial mode is expected to appear at steady state. 

\par
\smallskip
The following definition further classifies Turing instability %
into two types %
based on the location of the dominant pole. %
Note that the notation $\mathbb{N}$ does not include infinity.

\medskip
\noindent
{\bf Definition 3.~}
{\it 
The reaction-diffusion system with a single diffuser modeled by
(\ref{nonlinear-eq}) is Type-I Turing unstable around an equilibrium $\bar{{\bm
x}}$, if the following (A) and (B-I) are satisfied. Similarly, it is
Type-II Turing unstable if the following (A) and (B-II) are satisfied.

\smallskip
\noindent
(A) The system is Turing unstable.\\
(B-I) $p_k(s_0) = 0${\rm ;} $\exists k \in \mathbb{N}$ and $\exists s_0 \in \Pi$.\\
(B-II) $p_k(s_0) \neq 0${\rm ;} $\forall k \in \mathbb{N}$ and $\forall s_0 \in \Pi$.

\smallskip
\noindent
In other words, the system is Type-I Turing unstable, if the
dominant closed-loop pole is given by $\Sigma_k$ with some finite $k$, 
and it is Type-II Turing unstable, if the dominant closed-loop
pole is the pole of $\Sigma_k$ as $k \rightarrow \infty$.
}%

\begin{figure}[tb]
\centering
\includegraphics[clip,width=12cm]{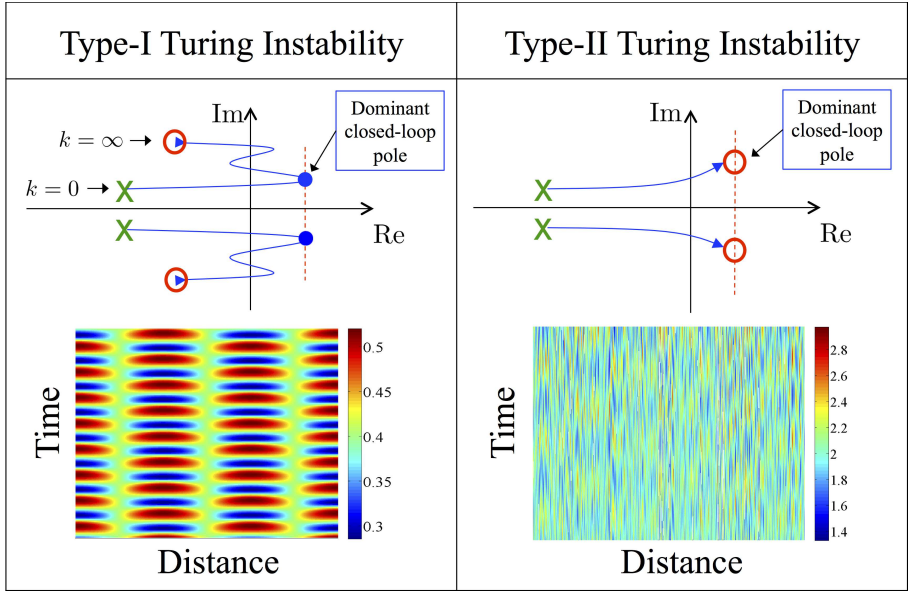}
\caption{Root loci and spatial patterns associated with Type-I and
 Type-II Turing instabilities.}
\label{types-fig}
\end{figure}

\medskip
Figure \ref{types-fig} illustrates the two types of Turing instability and
the corresponding spatio-temporal evolutions of the diffuser. 
In what follows, the spatial patterns associated with Type-I and
Type-II Turing instabilities are referred to as Type-I and Type-II Turing
patterns, respectively.

\par
\smallskip
We see from Fig. \ref{types-fig} that the dominant spatial mode is expected to be infinite for
Type-II Turing patterns, while it is finite for Type-I Turing patterns. 
The spatial patterns with the infinitely large spatial mode are, however,
questionable from a viewpoint of biological plausibility. 
Thus, we shall explore the conditions for Type-I Turing instability,
which excludes the biologically implausible patterns in Section V.
\medskip
\noindent
{\bf Remark 1.~}
Definitions 1,2 and 3 can also be used when there are multiple diffusers
in the system. For such cases, the characteristic polynomial $p_k(s)$
can be obtained by $p_k(s) = |sI - A + \lambda_k \hat{D}|$, 
where $\hat{D} := \mathrm{diag}(\mu_1, \mu_2, \cdots, \mu_n)$.
Nevertheless, the classification of Type-I and Type-II Turing instabilities
was not actively studied in the classical works of reaction-diffusion
systems \cite{Murray2003, Edelstein-Keshet2005} where the systems are composed
of $n=2$ molecules and both can diffuse in the spatial domain. 
This is because such a system is always Type-I Turing unstable, when it
is Turing unstable.
On the other hand, Type-II Turing instability can happen when the
number of diffusers is restricted. 
In \cite{Hsia2012}, it was shown that a reaction-diffusion system with one
diffuser can exhibit Turing instability, but the classification of the
instability was not discussed.
In fact, the examples presented in \cite{Hsia2012} are classified into
Type-II Turing instability. 
Hence, there still remains an important question that whether Type-I Turing
instability, which is physically plausible, is possible in
reaction-diffusion systems with a single diffuser.
$\hfill \Box$

\medskip
\section{Characterization of Turing Instabilities by Root Locus}

In this section, we show that the two types of Turing instability 
introduced in the previous section can be characterized by the root
locus of $h(s)$, which is the local reaction dynamics.
This characterization then leads to analytic conditions 
for Type-I Turing instability in the next section.

\par
\smallskip
It should be first noticed that $\Sigma_k$ is a negative feedback system
with the feedback gain $\lambda_k$ defined in (\ref{lambda-def}), and the feedback gain $\lambda_k$ monotonically
increases in terms of $k$. 
This implies that the poles of the subsystems
$\Sigma_k~(k=0,1,2,\cdots)$ lie on the root locus of the local reaction
dynamics $h(s)$. %
In other words, the poles of the subsystems $\Sigma_k~(k=0,1,2,\cdots)$ move from 
the poles toward the zeros of $h(s)$ as $k \rightarrow \infty$.
This allows us to characterize the two types of Turing instabilities
using the root locus of $h(s)$ as seen below.

\par
\smallskip
In order to examine the root locus in detail, we 
partition the matrix $A \in \mathbb{R}^{n \times n}$ as 
\begin{equation}
A = \left[
\begin{array}{c:c}
\tilde{A} & {\bm b} \\ \hdashline
{\bm c} & d
\end{array}
\right], 
\end{equation}
where $\tilde{A} \in \mathbb{R}^{(n-1) \times (n-1)}, {\bm b} \in
\mathbb{R}^{n-1}, {\bm c} \in \mathbb{R}^{1 \times (n-1)}$ and $d \in
\mathbb{R}$. 
The matrix $\tilde{A}$ represents the interactions between non-diffuser
molecules, and $d$ is the production or degradation rate of 
the single diffusible molecule. The vectors ${\bm b}$ and ${\bm c}$ 
express the interactions between the non-diffusers and the diffuser.
Then, the denominator $d(s)$ and the numerator $n(s)$ of $h(s)$ can be
specifically calculated as follows.

\medskip
\noindent
{\bf Lemma 1.~}
The local reaction dynamics $h(s)$ is written as 
\begin{align}
h(s) = \frac{n(s)}{d(s)} = %
 \frac{|sI_{n-1} - \tilde{A}|}{|sI_n - A|}.
\label{h-decomp-eq}
\end{align}

\medskip
\noindent
{\bf Proof.}
We first note that 
\begin{align}
h(s) = \frac{{\bm e_n}^T \mathrm{adj}(sI_n - A) {\bm e_n}}{|sI_n - A|}, 
\end{align}
where $\mathrm{adj}(X)$ represents the adjugate matrix of $X$.
It follows from the definition of the adjugate matrix that 
the $(n,n)$-th entry of $\mathrm{adj}(sI_n - A)$ is $|sI_{n-1} - \tilde{A}|$. 
Thus, ${\bm e}_n^T \mathrm{adj}(sI_n - A) {\bm e}_n = |sI_{n-1} - \tilde{A}|$.
$\hfill \Box$

\medskip
\noindent
{\bf Remark 2.~}
Lemma 1 provides an important physical interpretation of the denominator and the numerator
of $h(s)$. Specifically, we can see that that the zeros of $h(s)$ are
determined from the dynamics of the non-diffusers, $\mathcal{M}_1, \mathcal{M}_2,
\cdots, \mathcal{M}_{n-1}$, which is specified by $\tilde{A}$. 
The poles are, on the other hand, determined from the overall
local reaction dynamics including the diffuser $\mathcal{M}_n$.
This implies that the starting and the ending point of the root locus
can be characterized by these two dynamics.
$\hfill \Box$

\medskip
The following lemma summarizes the relation between the stability of 
subsystems $\Sigma_k~(k=0,1,2,\cdots)$ and the root locus of $h(s)$.

\medskip
\noindent
{\bf Lemma 2.~}
{\it 
The poles of $\Sigma_k~(i=1,2,\cdots)$ are given by the roots of 
\begin{align}
p(\lambda, s) = d(s) + \lambda n(s) = |sI_n - A| + \lambda |s I_{n-1} - \tilde{A}| = 0
\notag 
\end{align}
with $\lambda = \lambda_k$.
In particular, they are given by $\mathrm{spec}(A)$ for $k=0$ and move to
$\mathrm{spec}(\tilde{A})$ as $k \rightarrow \infty$.
}

\par
\medskip
We can see that that the difference between Type-I and Type-II Turing
instabilities essentially boils down to the relation between
the dominant closed-loop pole and the terminal point of the root locus
(see Fig. \ref{types-fig}). 
In particular, Lemma 2 implies that the terminal of the root locus is $\mathrm{spec}(\tilde{A})$. 
In the next section, these properties allow us 
to analytically explore conditions for Type-I Turing
instability.

\par
\smallskip
We note that the poles of $\Sigma_k~(k=0,1,2,\cdots)$ are discrete, 
whereas the root locus of $h(s)$ is continuous in terms of the feedback
gain $\lambda$. 
Thus, careful treatment is necessary to rigorously study instability
conditions based on the root locus. 
In practice, however, the length of the spatial domain $L$ is
sufficiently large such that the discrete feedback gains $\lambda_k = \mu(k\pi/L)^2~(k=0,1,2,\cdots)$
are close to each other. 
Moreover, our interest here is not the instability induced by the size
of the domain but the diffusion of a molecule. 
Hence, following the convention \cite{Murray2003, Edelstein-Keshet2005}, we hereafter analyze
Turing instability under the next assumption. 

\medskip
\noindent
{\bf Assumption 1.~}
{\it 
We assume $L$ is sufficiently large such that 
if $p(\lambda, s) = 0$ for some $\lambda > 0$ and some $s \in \mathbb{C}_+$, 
then $p(\lambda_k, s) = 0$ for some $k \in \mathbb{N}\cup\{\infty\}$ and some
$s \in \mathbb{C}_+$.
}

\medskip
\section{Conditions for Type-I Turing Instability}
In this section, we first provide a general necessary and sufficient condition for Type-I Turing
instability in reaction-diffusion systems with one diffuser. 
Using this condition, we further explore Type-I Turing instability for systems composed of $n=2$
and $n=3$ molecules.

\subsection{Necessary and sufficient condition}
We can first derive the following necessary and sufficient condition for
Type-I Turing instability from Lemma 2.

\medskip
\noindent
{\bf Lemma 3.~}
{\it 
Consider the reaction-diffusion system with a single diffuser modeled by
(\ref{nonlinear-eq}).
This system is Type-I Turing unstable around $\bar{{\bm x}}$, if and only if (i), and (ii-a) or
(ii-b) hold.

\smallskip
\noindent
{\rm (i)}~ $A$ is Hurwitz. 

\smallskip
\noindent
{\rm (ii-a)}~ $\tilde{A}$ is Hurwitz, and 
\begin{align}
\mathrm{Re}[p(\lambda, j \omega)] = 0,~~
\mathrm{Im}[p(\lambda, j \omega)] = 0
\end{align}
for some $\lambda > 0$ and some $\omega \in \mathbb{R}$.

\smallskip
\noindent
(ii-b)~ $\tilde{A}$ is not Hurwitz, and 
\begin{align}
\mathrm{Re}[p(\lambda, \beta + j \omega)] = 0,~~
\mathrm{Im}[p(\lambda, \beta + j \omega)] = 0
\end{align}
for some $\lambda > 0$ and some $\omega \in \mathbb{R}$, 
where 
$\beta := \max_{\nu \in \mathrm{spec}(\tilde{A})} \mathrm{Re}[\nu]$}

\par
\medskip
The conditions (ii-a) and (ii-b) guarantee that the rightmost zeros of 
the local reaction dynamics $h(s)$ are not identical with the dominant closed-loop pole, thus the
root locus is of Type-I in Fig. \ref{types-fig}. 
The difference between (ii-a) and (ii-b) is whether $h(s)$ is minimum
phase or not, %
that is, the terminal point of the root locus locates in the
left-half complex plane or not, respectively.
In the next section, we analytically derive conditions for Type-I Turing
instability based on Lemma 3.

\par
\smallskip
Using the properties of the root locus, we can also prove the following
proposition. %

\medskip
\noindent
{\bf Proposition 1.~}
{\it 
Consider the reaction-diffusion system with a single diffuser modeled by
(\ref{nonlinear-eq}). If the system is Type-I Turing unstable, 
the dominant closed-loop poles satisfy $\mathrm{Im}[\sigma] \neq 0$ for 
all $\sigma \in \Pi$.
}%

\par
\medskip
This proposition implies that the imaginary part of the dominant
closed-loop pole is always non-zero when the system has a single diffuser and is Type-I Turing unstable. Therefore, temporal oscillations are always expected together with
 spatial patterns.
\subsection{Conditions for systems composed of $n=2$ and $n=3$
	molecules}
In this section, we explore the question of whether Type-I Turing
instability is possible in reaction-diffusion systems with a single
diffuser, then we analytically derive conditions for Type-I Turing
instability. 
We define the coefficients $\alpha_i$ and $\tilde{\alpha}_i$ of the
characteristic polynomials as follows.
\begin{align}
&|sI_n - A| = s^n + \alpha_{n-1} s^{n-1} +  \cdots + \alpha_0, \\
&|sI_{n-1} - \tilde{A}| = s^{n-1} + \tilde{\alpha}_{n-2}s^{n-2} + \cdots
 + \tilde{\alpha}_0.
\end{align}

\par
\smallskip
Motivated by the classical activator-inhibitor model \cite{Gierer1972}, we 
first study the case of $n=2$. 

\medskip
\noindent
{\bf Theorem 1.~}
{\it 
Consider the reaction-diffusion system (\ref{nonlinear-eq}) with a
single diffuser composed of $n=2$ 
molecules. This system cannot be Type-I Turing unstable.
}

\medskip
\noindent
{\bf Proof.~}
We first note that $|sI - \tilde{A}| = s + \tilde{\alpha}_0$. 
In what follows, the proof will be given based on Lemma 3. 
Suppose $A$ is Hurwitz, which corresponds to the condition (i) in Lemma 
3, and we consider the following two cases.

\smallskip
\noindent
{\bf Case 1: $\tilde{A}$ is Hurwitz:}~
It follows from the definition that $\mathrm{Im}[p(\lambda, j\omega)] = \omega
(\alpha_1 + \lambda)$. 
We see that $\mathrm{Im}[p(\lambda, j\omega)] \neq 0$ for all $\lambda >
0$ and $\omega \neq 0$, since $\alpha_1 > 0$ holds when $A$ is Hurwtiz. 
For $\omega = 0$, it follows that $\mathrm{Re}[p(\lambda, 0)] =
\alpha_0 + \lambda \tilde{\alpha}_1$, and $\mathrm{Re}[p(\lambda,
j\omega)] \neq 0$ for all $\lambda > 0$.
Thus, the condition (ii-a) in Lemma 3 cannot be satisfied. 

\smallskip
\noindent
{\bf Case 2: $\tilde{A}$ is not Hurwitz:}~
It follows that $\mathrm{Im}[p(\lambda, \beta + j\omega)] =
\omega(2\beta + \alpha_1 + \lambda)$. Note that the definition of
$\beta$ implies $\beta = -\tilde{\alpha}_0$, and $\beta \ge 0$, because $\tilde{A}$ is not Hurwitz.
We see that $\mathrm{Im}[(\lambda, \beta + j\omega)] \neq 0$ for all
$\lambda > 0$ and $\omega \neq 0$, since $\beta \ge 0$ and $\alpha_1 >
0$. 
For $\omega = 0$, it follows that $\mathrm{Re}[p(\lambda, \beta)] =
\beta^2 + \alpha_0 + \lambda \tilde{\alpha}_1$, thus $\mathrm{Re}[p(\lambda,
j\omega)] \neq 0$ for all $\lambda > 0$.
This implies that the condition (ii-b) in Lemma 3 cannot be
satisfied. 

\par
\smallskip
Therefore, neither (ii-a) nor (ii-b) in Lemma 3 can be 
satisfied when $A$ is Hurwitz, which completes the proof.
$\hfill \Box$

\par
\medskip 

This theorem implies that biologically plausible patterns 
cannot be obtained by any interactions of $n=2$ molecules unlike the
activator-inhibitor models \cite{Gierer1972}.

\par
\smallskip
Thus, at least $n=3$ molecules are necessary to obtain 
Type-I Turing patterns. 
The following theorem provides analytic conditions for Type-I Turing
instability for the case of $n=3$.

\medskip
\noindent
{\bf Theorem 2.~}
{\it 
Consider the reaction-diffusion system (\ref{nonlinear-eq}) with a single
diffuser composed of $n=3$ molecules.
Then, the conditions (i), (ii-a) and (ii-b) in Lemma 3 are satisfied, if and only if 
the following (I), (II-A) and (II-B) are satisfied.
\begin{align}
\mathrm{(I)}&~~\alpha_2 > 0,~\alpha_1 \alpha_2 - \alpha_0 > 0,~\alpha_0 >
 0,\notag \\
\mathrm{(II \mathchar`- A)}&~~
\tilde{\alpha}_1 > 0,~\tilde{\alpha}_0 > 0, \notag \\
&~~\alpha_1 + \tilde{\alpha}_1 \alpha_2 - \tilde{\alpha}_0 \le 
-2 \sqrt{\tilde{\alpha}_1(\alpha_1 \alpha_2 - \alpha_0)}, \notag \\
\mathrm{(II \mathchar`- B)}&~~
\tilde{\alpha}_1 \le 0,~\tilde{\alpha}_1^2 - 4 \tilde{\alpha}_0 < 0,
 \notag \\ 
&~~-\tilde{\alpha}_1^2 + \tilde{\alpha}_0 + \tilde{\alpha}_1
 \alpha_2 - \alpha_1 > 0.
\notag
\end{align}
}%

\par
\medskip
In fact, the conditions (I), (II-A) and (II-b) in Theorem 2 
are equivalent to (i), (ii-a) and (ii-b) in Lemma 3, respectively.
The proof of Theorem 2 can be found in Appendix \ref{theo2-proof}.

\par
\smallskip
Using Theorems 2, we can determine the existence of Type-I Turing
patterns for a given reaction-diffusion system. 
In particular, the parameters $\tilde{\alpha}_i~(i=0,1)$ are obtained
only from the dynamics of non-diffuser molecules, {\it i.e.,} the matrix $\tilde{A}$.
Thus, given the dynamics of non-diffuser molecules, we can determine how
the dynamics of diffuser molecule and its interaction with 
non-diffusers, {\it i.e.,} $d, {\bm b}$ and ${\bm c}$, should be chosen
by tuning $\alpha_i~(i=0,1,2)$. 
\section{Numerical Example of Extended Gray-Scott Model}

\begin{figure}[tb]
\centering
\includegraphics[clip,width=8cm]{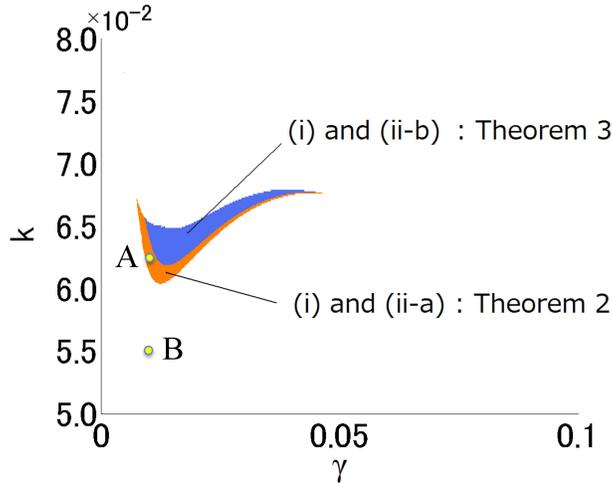}
\caption{The parameter region for Type-I Turing instability for the extended Gray-Scott
 model (\ref{gray-eq}). The marks $A$ and $B$ correspond to the
 parameter sets in Table \ref{param-table}.}
\label{param-fig}
\end{figure}

In this section, we consider the extended Gray-Scott model
\cite{Vastano1988} as an example of reaction-diffusion systems with $n=3$
molecules and demonstrate Type-I Turing patterns induced by a single diffuser.
Let the three molecules be denoted by $X, Y$ and $Z$. 
The local reactions of these molecules are written as 
\begin{align}
X + 2Y \rightleftarrows 3Y,~~~Y \rightleftarrows Z. \notag
\end{align}
We here assume that $X$ and $Y$ are non-diffusers and only $Z$ can diffuse in 
the spatial domain $\Omega=[0,1]$, although $Y$ and $Z$ were also assumed as diffusers
in the original model in \cite{Vastano1988}.

\begin{figure}[tb]
\centering
\includegraphics[clip,width=7.5cm]{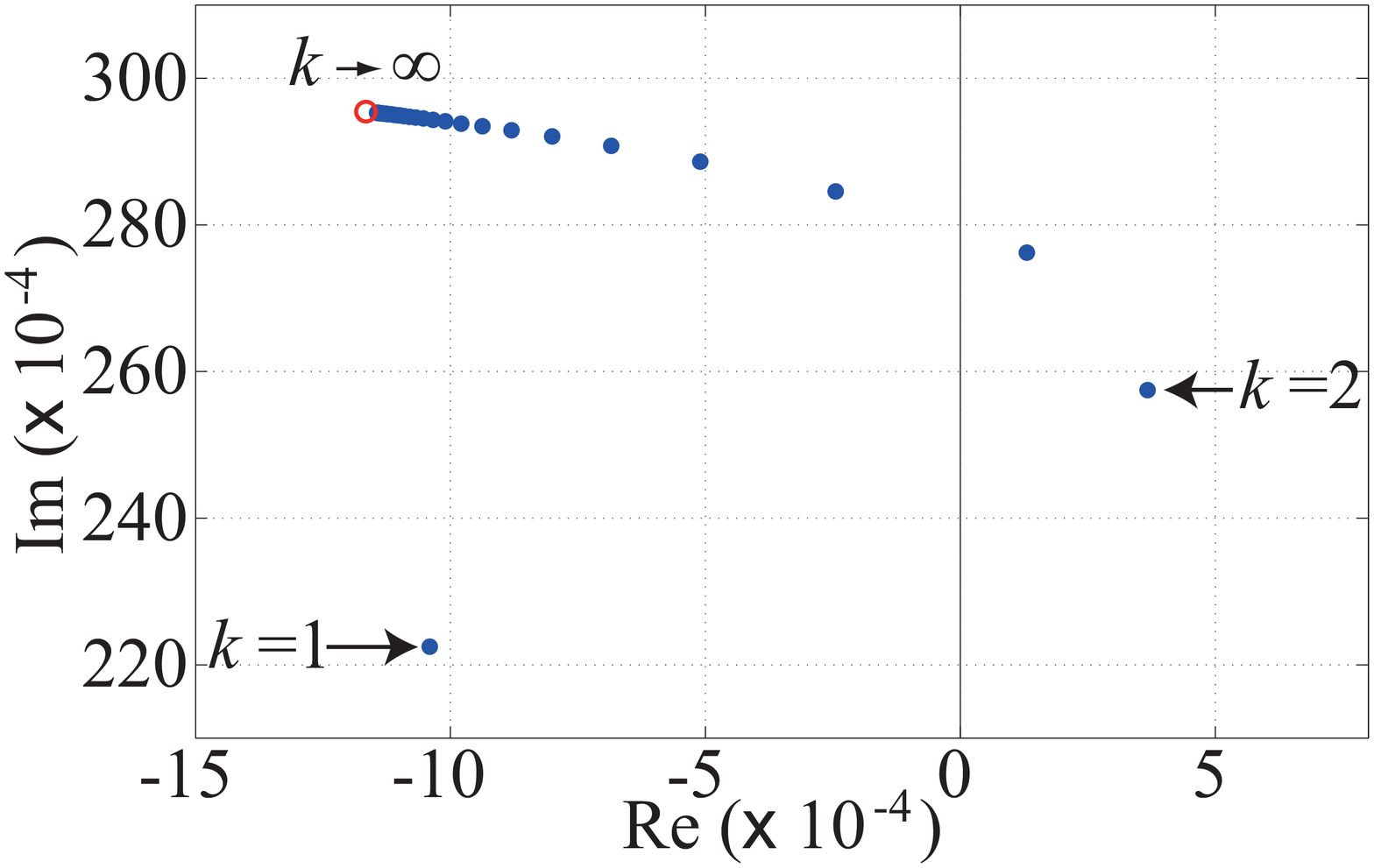}
\includegraphics[clip,width=6.0cm]{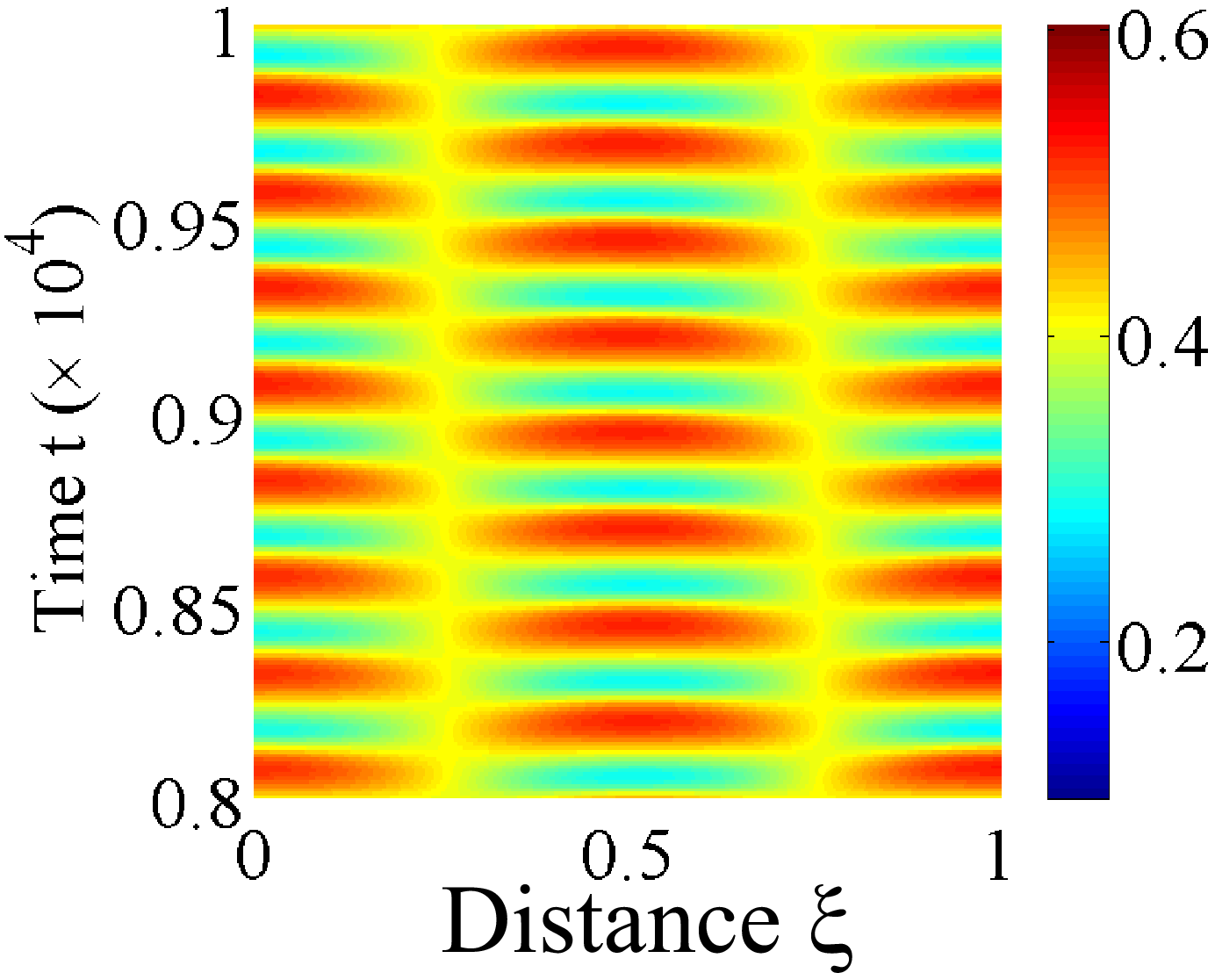}
\caption{(Left) Root locus of $h(s)$ for the parameter set A, 
which satisfies Theorem 2. In this example, $\tilde{A}$ is Hurwitz, and the dominant
 closed-loop pole is given by $\Sigma_2$. (Right) Spatio-temporal patterns 
of $X$ obtained by the simulation of the system (\ref{gray-eq}). The second
 spatial mode was observed.}
\label{sim-fig}
\end{figure}
\begin{figure}
\centering
\includegraphics[clip,width=7.5cm]{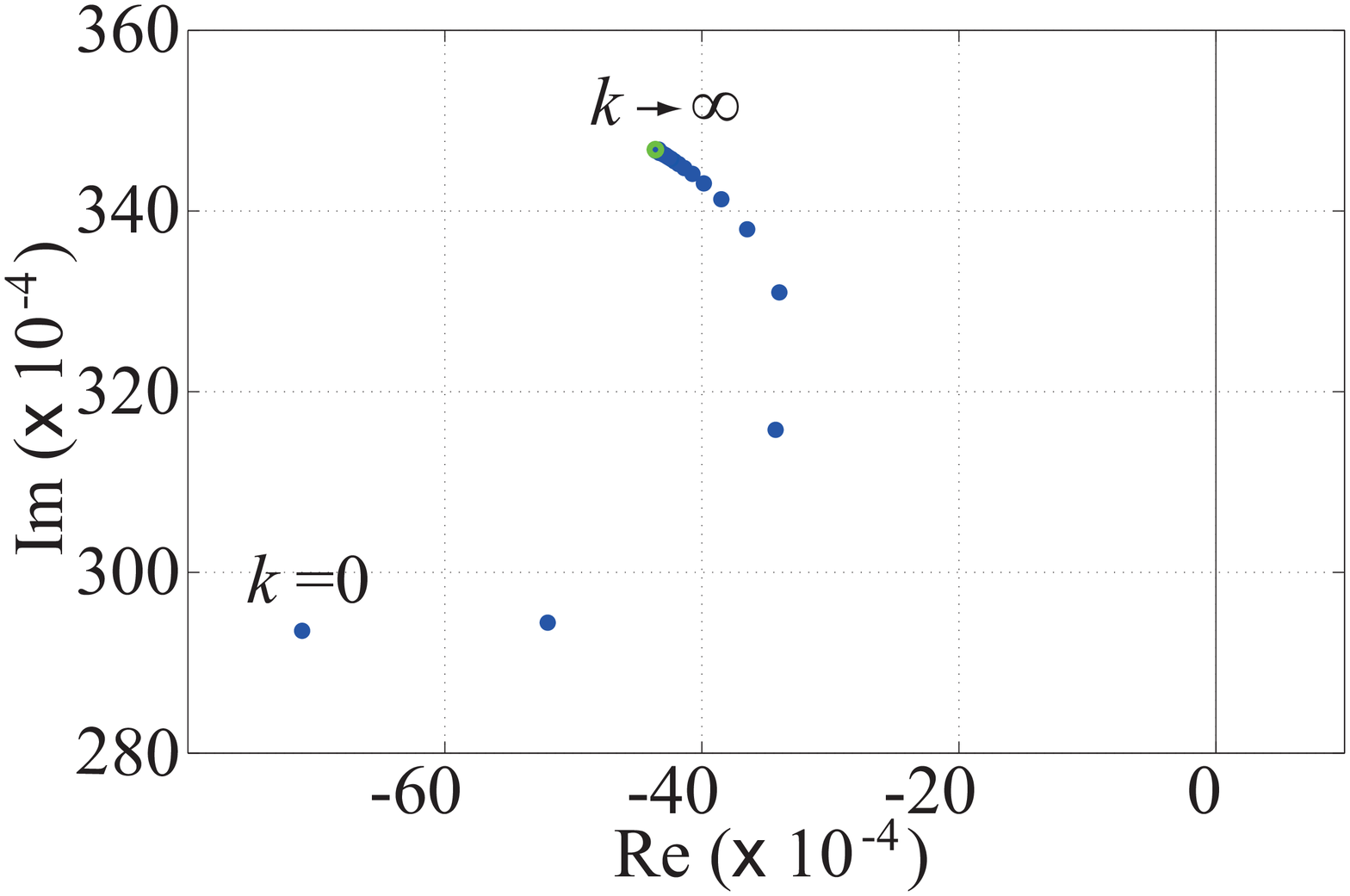}
\includegraphics[clip,width=6.2cm]{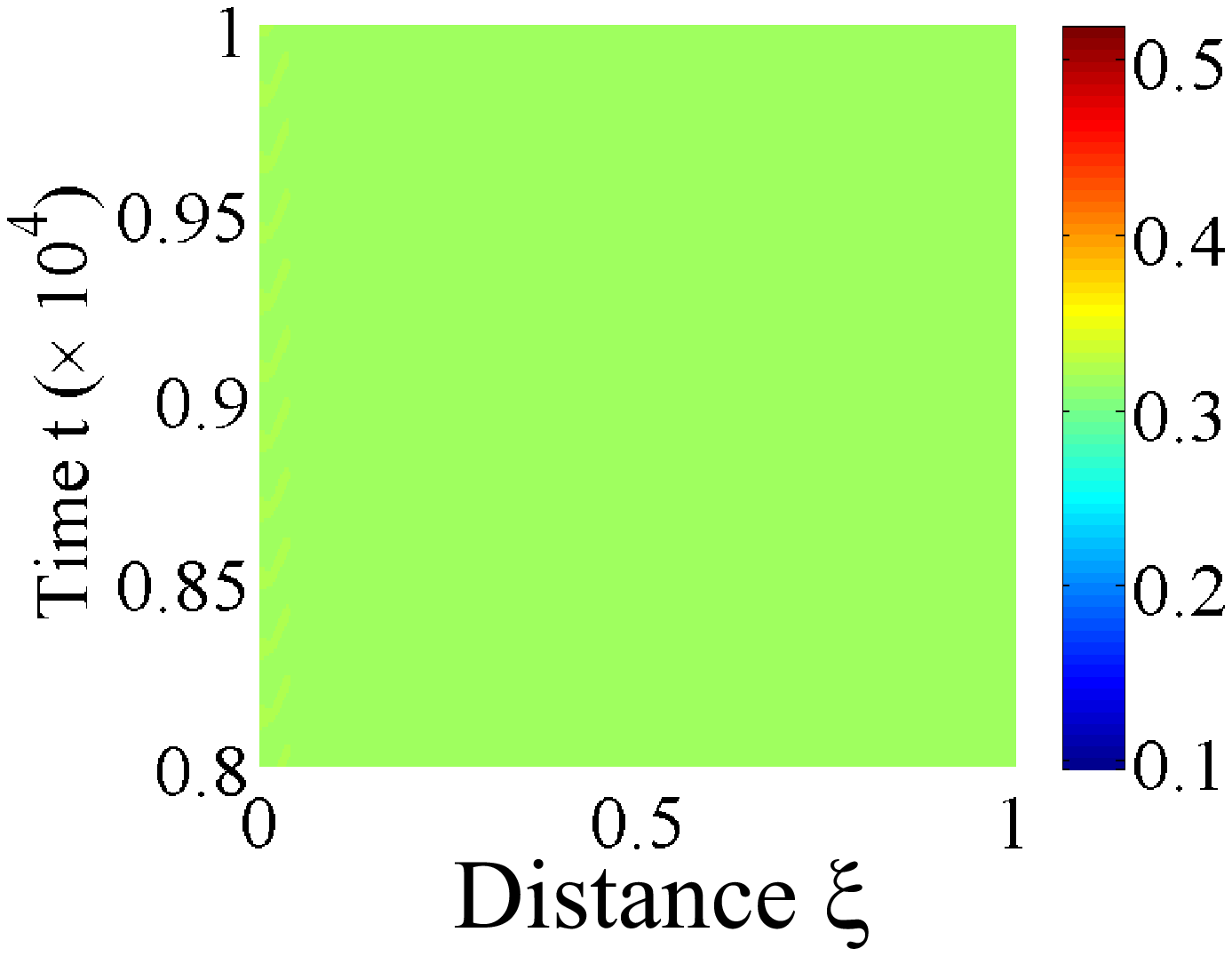}
\caption{(Left) Root locus of $h(s)$ for the parameter set B (see
 Fig. \ref{param-fig}). 
(Right) The concentration of $X$ simulated by (\ref{gray-eq}).}
\label{sim2-fig}
\end{figure}
\par
\smallskip
The dynamics of the extended Gray-Scott model with a single diffuser are
then obtained as 
\begin{align}
\begingroup
\renewcommand{\arraystretch}{2.0}
\begin{array}{ll}
\displaystyle
\frac{\partial x}{\partial t} = -x y^2 + \eta_1 y^3 + \gamma (1 - x) \\
\displaystyle\frac{\partial y}{\partial t} = xy^2 - \eta_1 y^3 - k (y - \eta_2 z)
 - \gamma y \\
\displaystyle
\frac{\partial z}{\partial t} = k (y - \eta_2 z) - \gamma z + \mu
 \nabla^2 z,
\end{array}
\endgroup
\label{gray-eq}
\end{align}
where $x(\xi, t), y(\xi, t)$ and $z(\xi, t)$ are normalized
concentrations of $X, Y$ and $Z$, respectively. 
The parameters $(\eta_1, \eta_2, k, \gamma)$ and $\mu$ are normalized
reaction and diffusion rates, respectively (see \cite{Vastano1988} for the details).

\par
\smallskip
In what follows, we linearize the system (\ref{gray-eq}) around a
homogeneous equilibrium and demonstrate a Type-I Turing pattern
predicted by Theorem 2. 
To this end, we first obtain the three homogeneous equilibria of
(\ref{gray-eq}) as 
\begin{align}
&(x_0, y_0, z_0) = (1,0,0), \notag \\
&(x_+, y_+, z_+) =\left(
\frac{\eta_1 y_+^3 + \gamma}{y_+^2 + \gamma}, 
\frac{v + \sqrt{w}}{2\left((1 + \eta_1)v + k\right)}, 
\frac{k y_+}{v}
\right), \notag \\
&(x_-, y_-, z_-) = \left(
\frac{\eta_1 y_-^3 + \gamma}{y_-^2 + \gamma}, 
\frac{v - \sqrt{w}}{2\left((1 + \eta_1)v + k\right)}, 
\frac{k y_-}{v}
\right), \notag
\end{align}
where $w := v^2 - 4\gamma (k + v)\left\{(1 + \eta_1)v + k\right\}$
 and $v:=\gamma + k \eta_2$.
The Jacobian of (\ref{gray-eq}) at $(x_*, y_*, z_*)$ is then calculated as 
\begin{align}
\begin{bmatrix}
-y_*^2 - \gamma & -2x_* y_*  + 3\eta_1 y_*^2 & 0 \\
y_*^2 & 2x_* y_* - 3\eta_1 y_*^2 - k -\gamma & k \eta_2 \\
0 & k & -k \eta_2 - \gamma
\end{bmatrix}. 
\end{align}
Note that we can verify that the equilibrium $(x_+, y_+, z_+)$ can be Turing unstable,
whereas $(x_0, y_0, z_0)$ is always stable.

\par
\smallskip
Hence, we hereafter explore Type-I Turing instability around the
equilibrium $(x_+, y_+, z_+)$.
Let $\eta_1 = \eta_2 = 0.1$. The Jacobian then depends on the two parameters $\gamma$ and $k$, which
are the normalized degradation and production rates of the diffuser $Z$,
respectively. 
Thus, using Theorem 2, we can specify the parameter region of 
$(\gamma, k)$ such that the system (\ref{gray-eq}) is Type-I Turing unstable around
$(x_+, y_+, z_+)$ as illustrated in Fig. \ref{param-fig}.

\par
\smallskip
In order to verify the Type-I Turing unstable condition, we numerically
simulated the model (\ref{gray-eq}) for the parameter sets shown in
Table \ref{param-table}.
\begin{table}[tb]
\centering
\caption{The parameter sets for the simulations in Fig. \ref{sim-fig}}
\label{param-table}
\begin{tabular}{c|ccc} \hline \hline
Set & $\gamma$ & $k$ & $\mu$ \\ \hline
A & $1.0 \times 10^{-2}$ & $6.2 \times 10^{-2}$ & $1.0 \times 10^{-3}$ \\
B & $1.0 \times 10^{-2}$ & $5.5 \times 10^{-2}$ & $1.0 \times 10^{-3}$ \\ \hline
\end{tabular}
\end{table}
The simulation results are illustrated in Fig. \ref{sim-fig} (Right) and
Fig. \ref{sim2-fig} (Right) for the sets $A$ and $B$, respectively
We can verify that the chemical concentration forms the Type-I Turing
pattern when the parameter is chosen from the 
colored region in Fig. \ref{param-fig}, while it converges to a spatially
homogeneous equilibrium point for the parameters outside the region. 
Moreover, we can see that the profile of the spatial pattern in
Fig. \ref{sim-fig} (Right) is the second mode corresponds to the fact that the dominant
closed-loop pole is given by $\Sigma_2$ (see Fig. \ref{sim-fig} (Left)).

\par
\smallskip
The simulations in Figs. \ref{sim-fig} and \ref{sim2-fig} were implemented by the command `{\tt pdepe}'
 in MATLAB R2010b. 
The initial conditions were set as 
$[x(\xi,0), y(\xi,0), z(\xi,0)]^T = \sum_{k=1}^{20}0.01$ 
$\cos(k\pi
\xi)[1,1,1]^T$ for both examples.
\medskip
\section{Conclusion}
In this paper, we have developed a control theoretic framework to
analyze the reaction-diffusion systems with a single diffusible
component. 
We have first shown that the instability analysis of the
reaction-diffusion system can be greatly simplified by formulating  
as a multi-agent dynamical system.
This formulation has then allowed us to characterize biologically plausible and implausible
Turing instabilities using the root locus of each agent's dynamics.
Thus, using the properties of the root locus, 
we have proven that the reaction-diffusion
systems can exhibit biologically plausible Turing patterns even when 
there is only one diffusible component.
Then, the conditions for the pattern formation have been analytically
obtained for systems with three components.

\medskip
\noindent
{\bf Acknowledgments: }
This work was supported in part by the Ministry of Education,
Culture, Sports, Science and Technology in Japan through 
Grant-in-Aid for Scientific Research (A) No. 21246067, and 
Grant-in-Aid for JSPS Fellows No. 23-9203.

\bibliographystyle{plain}
\bibliography{20130313}

\begin{thebibliography}{10}

\bibitem{Basu2005}
S.~Basu, Y.~Gerchman, C.~H. Collins, F.~H. Arnold, and R.~Weiss.
\newblock A synthetic multicellular system for programmed pattern formation.
\newblock {\em Nature}, 434(7037):1130--1134, 2005.

\bibitem{Basu2004}
S.~Basu, R.~Mehreja, S.~Thiberge, M.-T. Chen, and R.~Weiss.
\newblock Spatiotemporal control of gene expression with pulse-generating
  networks.
\newblock {\em Proceedings of National Academy of Sciences},
  101(17):6355--6360, 2004.

\bibitem{Curtain1995}
R.~F. Curtain and H.~J. Swart.
\newblock {\em An introduction to infinite-dimensional linear systems theory}.
\newblock Springer, 1995.

\bibitem{Edelstein-Keshet2005}
L.~Edelstein-Keshet.
\newblock {\em Mathematical models in biology}.
\newblock SIAM, 2005.

\bibitem{Gierer1972}
A.~Gierer and H.~Meinhardt.
\newblock A theory of biological pattern formation.
\newblock {\em Kybernetik}, 12(1):30--39, 1972.

\bibitem{Hara2013}
S.~Hara, H.~Tanaka, and T.~Iwasaki.
\newblock Stability analysis of systems with generalized frequency variables.
\newblock {\em IEEE Transactions on Automatic Control}.
\newblock (to appear).

\bibitem{Hsia2012}
J.~Hsia, W.~J. Holtz, D.~C. Huang, M.~Arcak, and M.~M. Maharbiz.
\newblock A feedback quenched oscillator produces {Turing} patterning with one
  diffuser.
\newblock {\em PLOS Computational Biology}, 8(1), 2012.
\newblock e1002331.

\bibitem{Koch1994}
A.J. Koch and H.~Meinhardt.
\newblock Biological pattern formation: from basic mechanisms to complex
  structures.
\newblock {\em Reviews of Modern Physics}, 66(4):1481--1507, 1994.

\bibitem{Kondo2010}
S.~Kondo and T.~Miura.
\newblock Reaction-diffusion model as a framework for understanding biological
  pattern formation.
\newblock {\em Science}, 329(5999):1616--1620, 2010.

\bibitem{Loose2011}
M.~Loose, K.~Kruse, and P.~Schwille.
\newblock Protein self-organization: lessons from the min system.
\newblock {\em Annual Review of Biophysics}, 40:315--336, 2011.

\bibitem{Miller2001}
M.B. Miller and B.L. Bassler.
\newblock Quorum sensing in bacteria.
\newblock {\em Annual Review of Microbiology}, 55:165--199, 2001.

\bibitem{Murray2003}
J.D. Murray.
\newblock {\em Mathematical Biology II}.
\newblock Springer, 3rd edition edition, 2003.

\bibitem{Tanaka2009}
H.~Tanaka, S.~Hara, and T.~Iwasaki.
\newblock {LMI} stability condition for linear systems with generalized
  frequency variables.
\newblock In {\em Proceedings of Asian Control Conference}, pages 136--141,
  2009.

\bibitem{Turing1952}
A.~M. Turing.
\newblock The chemical basis of morphogenesis.
\newblock {\em Philosophicals Transactions of the Royal Society of London B},
  237(641):37--72, 1952.

\bibitem{Vastano1988}
J.~A. Vastano, J.~E. Pearson, W.~Horsthemke, and H.~L. Swinney.
\newblock Turing patterns in an open reactor.
\newblock {\em The Journal of Chemical Physics}, 88(6175), 1988.

\end{thebibliography}

\appendix
\section{Proof of Theorem 2}
\label{theo2-proof}
We prove the theorem by showing equivalent conditions to (i), (ii-a) and 
(ii-b) in Lemma 3.

\par
\smallskip
It can be verified that $A$ is Hurwitz if and only if 
\begin{align}
\alpha_2 > 0,~\alpha_1 \alpha_2 - \alpha_0 > 0,~\alpha_0 > 0, 
\label{case1-A-eq}
\end{align}
which is equivalent to the condition (i) in Lemma 3. 

\par
\smallskip
In what follows, we derive the conditions (II-A) and (II-B) by considering the case where $\tilde{A}$ is Hurwitz and
not Hurwitz, respectively.
We can easily verify that neither (ii-a) nor (ii-b) of Lemma 3
can be satisfied when $\omega = 0$. 
Hence, we hereafter show the proof for the case of $\omega \neq 0$.

\medskip
\noindent
{\bf Case 1: $\tilde{A}$ is Hurwitz:~}
A necessary condition for $\tilde{A}$ being
Hurwitz is 
\begin{align}
\tilde{\alpha}_1 > 0~\mathrm{and}~\tilde{\alpha}_0 > 0.
\label{case1-tildeA-eq}
\end{align}
Moreover, $p(\lambda, j\omega) = 0$ implies that 
\begin{align}
&\mathrm{Re}[p(\lambda, j\omega)] = -\alpha_2 \omega^2 + \alpha_0 +
 \lambda(-\omega^2 + \tilde{\alpha}_0) = 0 \label{case1-real-eq}\\
&\mathrm{Im}[p(\lambda, j\omega)] = -\omega\{\omega^2 - (\alpha_1 +
 \lambda \tilde{\alpha}_1)\} = 0.\label{case1-imag-eq}
\end{align}
It follows from (\ref{case1-imag-eq}) that $\omega^2 = \alpha_1 +
\lambda \tilde{\alpha}_1 (> 0)$, since $\omega \neq 0$. 
We can then eliminate $\omega^2$ from (\ref{case1-real-eq}), and we have
the following quadratic equation of $\lambda$.
\begin{align}
\tilde{\alpha}_1 \lambda^2 + (\alpha_1 + \tilde{\alpha}_1 \alpha_2 -
 \tilde{\alpha}_0)
\lambda + \alpha_1 \alpha_2 - \alpha_0 = 0.
\label{case1-lambda-eq}
\end{align}
Since $\tilde{\alpha}_1 > 0$ and $\alpha_1 \alpha_2 - \alpha_0 > 0$
hold from (\ref{case1-A-eq}) and (\ref{case1-tildeA-eq}), a necessary
and sufficient condition for (\ref{case1-lambda-eq}) having a positive
real solution is given by the following (C1) and (C2).

\smallskip
\noindent
(C1) The determinant of (\ref{case1-lambda-eq}) is non-negative. That
is, 
\begin{align}
(\alpha_1 + \tilde{\alpha_1} \alpha_2 - \tilde{\alpha}_0^2)^2 -
 4\tilde{\alpha}_1(\alpha_1 \alpha_2 - \alpha_0) \ge 0.
\label{C1-eq}
\end{align}

\smallskip
\noindent
(C2) The coefficient of $\lambda$ in (\ref{case1-lambda-eq}) is
negative. That is, 
\begin{align}
\alpha_1 + \tilde{\alpha}_1 \alpha_2 - \tilde{\alpha}_0 < 0.
\label{C2-eq}
\end{align}

\par
\smallskip
Summarizing (\ref{C1-eq}) and (\ref{C2-eq}), we have 
\begin{align}
\alpha_1 + \tilde{\alpha}_1 \alpha_2 - \tilde{\alpha}_0 \le 
-2 \sqrt{\tilde{\alpha}_1 (\alpha_1 \alpha_2 - \alpha_0)}.
\label{case1-last-eq}
\end{align}
The condition (II-A) is obtained from (\ref{case1-tildeA-eq}) and 
(\ref{case1-last-eq}).

\medskip
\noindent
{\bf Case 2: $\tilde{A}$ is not Hurwitz~}
A necessary condition for $\tilde{A}$ not being Hurwitz is 
\begin{align}
\tilde{\alpha}_1 \le 0~\mathrm{or}~\tilde{\alpha}_0 \le 0.
\label{case2-tildeA-eq}
\end{align}
Let $\beta \pm j \gamma~(\beta, \gamma \ge 0)$ denote a pair of 
eigenvalues of $\tilde{A}$ with the largest real part. 
Then, $|(\beta + j\gamma)I_2 - \tilde{A}| = 0$ implies 
\begin{align}
&\beta^2 + \tilde{\alpha}_1 \beta + \tilde{\alpha}_0 - \gamma^2 = 0
\label{case2-tildeA-det-eq1}
\\
&(2 \beta + \tilde{\alpha}_1) \gamma = 0.
\label{case2-tildeA-det-eq2}
\end{align}

\par
\smallskip
In what follows, we first show that $p(\lambda, \beta + j\gamma) \neq 0$ for all $\lambda > 0$ 
and $\omega$ when $\gamma = 0$.
It follows from $\mathrm{Im}[p(\lambda, \beta + j\omega)] = 0$ that 
\begin{align}
\omega^2 = 3 \beta^2 + 2 \alpha_2 \beta + \alpha_1 + \lambda(2\beta +
 \tilde{\alpha}_1).
\end{align}
Substituting this into $\mathrm{Re}[p(\lambda, \beta + j\omega)] = 0$
yields the following equation of $\lambda$.
\begin{align}
(2\beta + \tilde{\alpha}_1) \lambda^2 &+ \{
9\beta^2 + (4 \alpha_2 + 3 \tilde{\alpha}_1)\beta + \alpha_1 +
 \tilde{\alpha}_1 \alpha_2
\}\lambda \notag \\
&+
\{
8\beta^3 + 8\alpha_2 \beta^2 + 2(\alpha_1 + \alpha_2^2)\beta
 + (\alpha_1 \alpha_2 - \alpha_0)
\} = 0.
\label{case2-gamma0-eq}
\end{align}
It should be noted that the coefficient of $\lambda^2$ is non-negative, or 
$2 \beta + \tilde{\alpha}_1 \ge 0$, since $\beta$ is the 
largest real root of $s^2 + \tilde{\alpha}_1 s + \tilde{\alpha}_0 = 0$
from the definition.
Moreover, (\ref{case1-A-eq}) and (\ref{case2-tildeA-eq}) imply that the 
coefficient of $\lambda$ and the constant terms of (\ref{case2-gamma0-eq}) 
are positive as well.
Thus, $p(\lambda, \beta + j\gamma) \neq 0$ for all $\lambda > 0$ 
and $\omega$ when $\gamma = 0$.

\par
\smallskip
When $\gamma \neq 0$, the determinant of $|sI_2 - \tilde{A}| = 0$ is
negative, thus we have 
\begin{align}
\tilde{\alpha}_1^2 - 4 \tilde{\alpha}_0 < 0.
\label{case2-det-eq}
\end{align}
Moreover, $\mathrm{Im}[p(\lambda, \beta + j\omega)] = 0$ implies 
\begin{align}
\omega^2 = 3 \beta^2 + 2\alpha_2 \beta + \alpha_1.
\end{align}
Substituting this into $\mathrm{Re}[p(\lambda, \beta + j\omega)] = 0$ yields
\begin{align}
\{\gamma^2 - (3\beta^2 + 2 \alpha_2 \beta + \alpha_1)\}\lambda 
 - 
\{
8\beta^3 + 8\alpha_2 \beta^2 + 2(\alpha_1 + \alpha_2^2)\beta + 
(\alpha_1 \alpha_2 - \alpha_0)
\}
 = 0.
\label{case2-gammanon0-eq}
\end{align}
Note that the constant term of (\ref{case2-gammanon0-eq}) is negative 
 because of (\ref{case1-A-eq}) and $\beta > 0$.
Thus, (\ref{case2-gammanon0-eq}) has a positive real root if and only if 
\begin{align}
\gamma^2 - (3\beta^2 + 2 \alpha_2 \beta + \alpha_1) > 0.
\label{case2-last-ineq}
\end{align}
The condition (II-B) is obtained from (\ref{case2-tildeA-eq}), 
(\ref{case2-tildeA-det-eq1}), (\ref{case2-tildeA-det-eq2}), 
(\ref{case2-det-eq}) and (\ref{case2-last-ineq}).
$\hfill \Box$

\end{document}